\documentclass[twocolumn,superscriptaddress,aps,preprintnumbers,amsmath,amssymb,prl,nofootinbib]{revtex4}

\usepackage{graphicx}
\usepackage{epstopdf}
\usepackage{dcolumn}
\usepackage{bm}
\usepackage{hyperref}
\usepackage{color}
\usepackage{amsmath}

\begin{document}

\title{
An onset model of mutually catalytic self-replicative systems formed by an assembly of polynucleotides
}

\author{Yasuji Sawada}
\affiliation{Division for Interdisciplinary Advanced Research and Education, Tohoku University, Sendai 980-8578, Japan}
\affiliation{Frontier Research Institute for Interdisciplinary Sciences, Tohoku University, Sendai 980-8578, Japan}
\author{Yasukazu Daigaku}
\affiliation{Frontier Research Institute for Interdisciplinary Sciences, Tohoku University, Sendai 980-8578, Japan}
\affiliation{Cancer Genome Dynamics project, Cancer Institute, Japanese Foundation for Cancer Research, Tokyo, Japan}
\author{Kenji Toma}
\affiliation{Frontier Research Institute for Interdisciplinary Sciences, Tohoku University, Sendai 980-8578, Japan}
\affiliation{Astronomical Institute, Graduate School of Science, Tohoku University, Sendai 980-8578, Japan}

\begin{abstract}
  Self-replicability is the unique attribute observed in all the living organisms and the question how the life was physically initiated could be equivalent to the question how self-replicating informative polymers were formed in the abiotic material world. It has been suggested that the present DNA and proteins world was preceded by RNA world in which genetic information of RNA molecules was replicated by the mutual catalytic function of RNA molecules. However, the important question how the transition occurred from a material world to the very early pre-RNA world remains unsolved experimentally nor theoretically. We present an onset model of mutually catalytic self-replicative systems formed in an assembly of polynucleotides. A quantitative expression of the critical condition for the onset of growing fluctuation towards self-replication in this model is obtained by analytical and numerical calculations.
\end{abstract}
\maketitle

%
%
%

\section{I. Introduction}

It has been widely suggested that the present DNA and proteins world was preceded by RNA world in which the genetic information resided in the sequence of RNA molecules and was copied by the mutual catalytic function of RNA molecules \cite{gilbert86,cech86,urau98,johnston01,robertson11,szostak16,lincoln09}.
The research of RNA world during the first period was mainly focused on the possibility of finding examples of self-replication even the nucleotide is short \cite{naylor66,orgel74,inoue83}. Inoue \& Orgel (1983) \cite{inoue83} reported first observation of a template with defined sequence can catalyze the formation of its complementary strand. Kiedrowski (1986) \cite{kiedrowski86} demonstrated a repeated ligation of short nucleotide chains including the separation of the complement from its template without the help of any enzyme.
Since the discovery of ribozyme \cite{cech86}, the research of RNA world was gradually directed to find  functions of various ribozymes for the self-replication process in the laboratory experiments \cite{lincoln09,pressman15}, and introduced various mechanisms, such as ligation \cite{joice02,paul02}, RNA polymerase ribozymes \cite{horning16}, strand-displacement \cite{zhou19}.

However, important questions of how non-enzymatically replication cycle started in the pre-RNA world remains unsolved \cite{robertson11,zhou19}. Transition from the material world to the beginning of RNA world has not been clarified experimentally nor theoretically. Difficulty in answering this problem by laboratory experiments may be partly due to the fact that the range of the experimental conditions are limited compared to that provided by the nature in the long time span of $10^9$ years of pre-RNA world. Therefore, the experimental results to date have not yet encompassed the model which correspond to the very beginning of self-replication mechanism.

Alternative approach to investigate the beginning of self-replication is to use theoretical one. There have been theoretical studies on the birth of life based on auto-catalytic cell model \cite{kauffman93,kauffman86,mossel05,hordijk15,hordijk17,hordijk18}, hypercycle model \cite{eigen71,eigen79,szostak16b,boerlijst91,sardanyes06} and chemical evolution \cite{higgs15,higgs17}. More recently, theoretical studies of rolling circle and strand-displacement mechanisms \cite{tupper21} or cooperative ligation mechanism for non-enzymatic self-replication \cite{toyabe19} were reported. However, these theoretical studies have not focused on the onset of self-replicability in the pre-RNA world, while they were either generalized to wider topics including evolution mechanism of Darwinian world \cite{darwin59} or characterized to more specific functions of some RNA molecules. The existing theories of pre-RNA world were based on the catalytic functions of various kinds, for example, self-learning catalyzers in case of hypercycle theories and ligases in case of autocatalytic theories. The reason for it is probably because these theories have been intended to build theories which covers the beginning and evolution of the RNA world, but not to clarify the transition itself from the material world.

The theoretical model of the present paper, on the other hand, was intended to describe the transition of a material world to the beginning of pre-RNA world in terms of the material world, without using any functional molecules which had not existed in the material world before the transition. For the present purpose the model is required to describe mathematically the occurrence of the transition at some parameter values of the well-defined material world. By the material world is meant here an interacting energy rich mononucleotide molecules of abundant density and gradually increasing poly-nucleotide molecules only. By pre-RNA world is meant a world with self-replicating poly-nucleotide molecules which is indispensable for the robust members of good information quality for the birth of life.

In Sec.~II, dynamical onset models of mutually catalytic self-replication are presented. In Sec.~III, two kinds of possible networks consisting of mutually catalytic polynucleotides are proposed. In Sections~IV and V, analyses and numerical simulations of the networks are presented, respectively. In Sec.~VI, critical conditions for the onset of self-replicator networks, scenario of realization of the condition, and its thermodynamic aspects are discussed.

\section{II. Dynamical onset models of mutually catalytic self-replicative systems}

\subsection{A. Material world just before the transition to an early RNA world}

It is believed that there must have been mono-nucleotide soup \cite{gilbert86,cech86,urau98,johnston01,robertson11,szostak16,lincoln09,naylor66} in some favorite locations of the earth of prebiotic worlds, which contained various building blocks of mononucleotide molecules (abbreviated hereafter as mn-molecules), such as sugars, phosphates, organic bases which served as law materials.
  
\begin{figure}
\begin{center}
  \includegraphics[height=6.0cm,keepaspectratio]{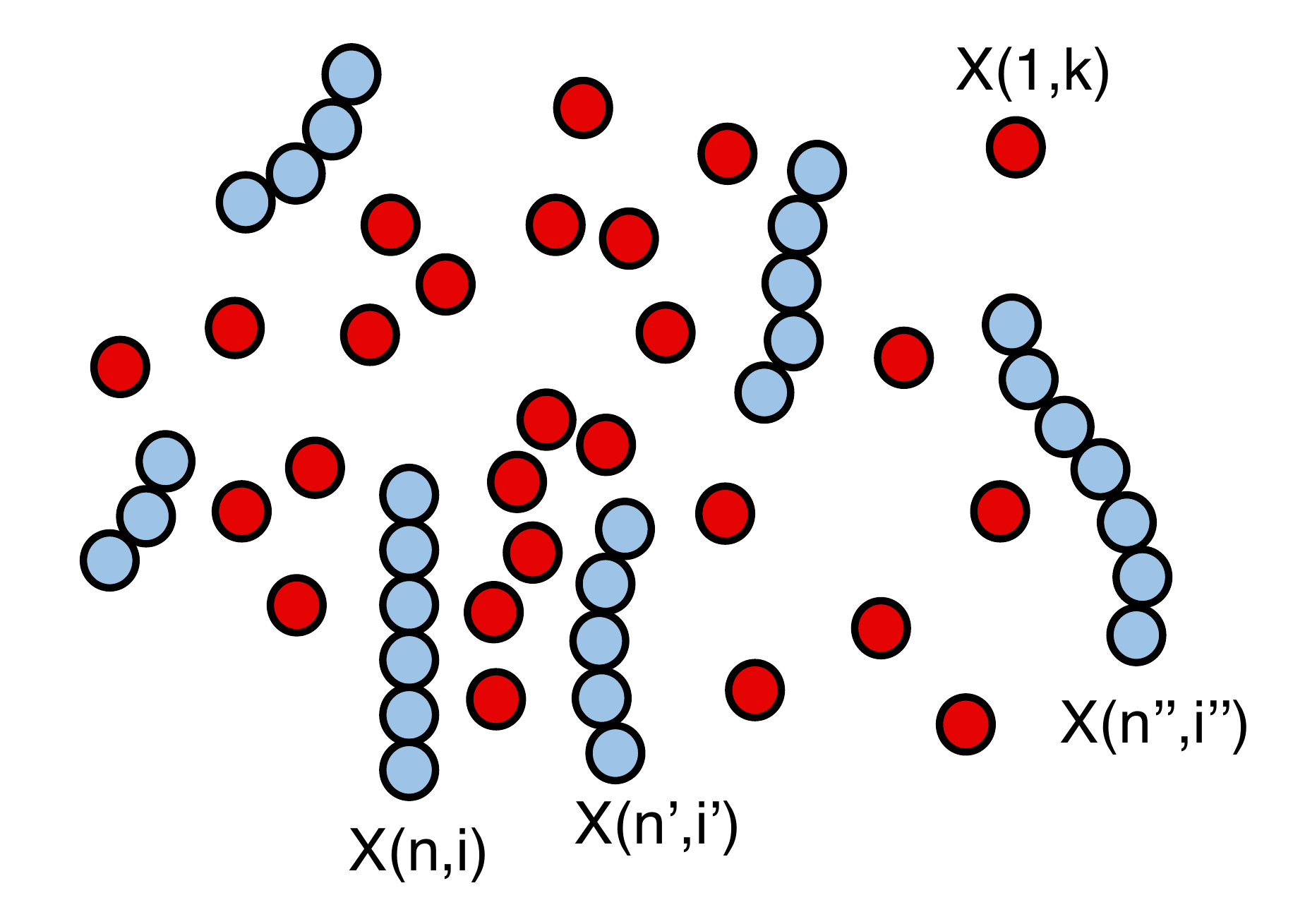}
  \caption{A pn-molecule $X(n,i)$ under consideration is surrounded by environmental pn-molecules $X(n',i')$'s and mn-molecules $X(1,k)$'s in the material world. The pn-molecules and mn-molecules are represented by the blue and red colors, respectively. One may assume that there is one pn-molecule $X(n',i')$, for example, which interacts most strongly with $X(n,i)$ under consideration.}
\label{fig:model}
\end{center}
\end{figure}

It would be natural to assume that in the long history of prebiotic period, the four kinds of mn-molecule $X(1,k) ~~(k=1,2,3,4)$ as well as various poly-nucleotide molecule $X(n,i)$ (abbreviated hereafter as pn-molecules) had accumulated in the material world before the transition. Here, $n$ represents the length, and $i$ is $i$-th order of nucleosides of a pn-molecule of length $n$. The density of the pn-molecules kept increasing gradually by the high energy mn-molecules. A pn-molecule $X(n,i)$ is surrounded by mn-molecules $X(1,k)$ and other pn-molecules $X(n’,i’)$, as shown in Fig.~\ref{fig:model}.  A mn-molecule $X(1,k)$ can stick to the nucleotide which is compliment to $k$-th mn-molecule in the pn-molecule $X(n,i)$. Although some double strands may have been formed by the spontaneous processes even in this period of material world, the density of the double strands did not increase enough due to possible decay process when $X(n,i)$'s are small.

\subsection{B. Dynamical onset model of mutually catalytic self-replication}

\begin{figure*}
\begin{center}
  \includegraphics[height=8.0cm,keepaspectratio]{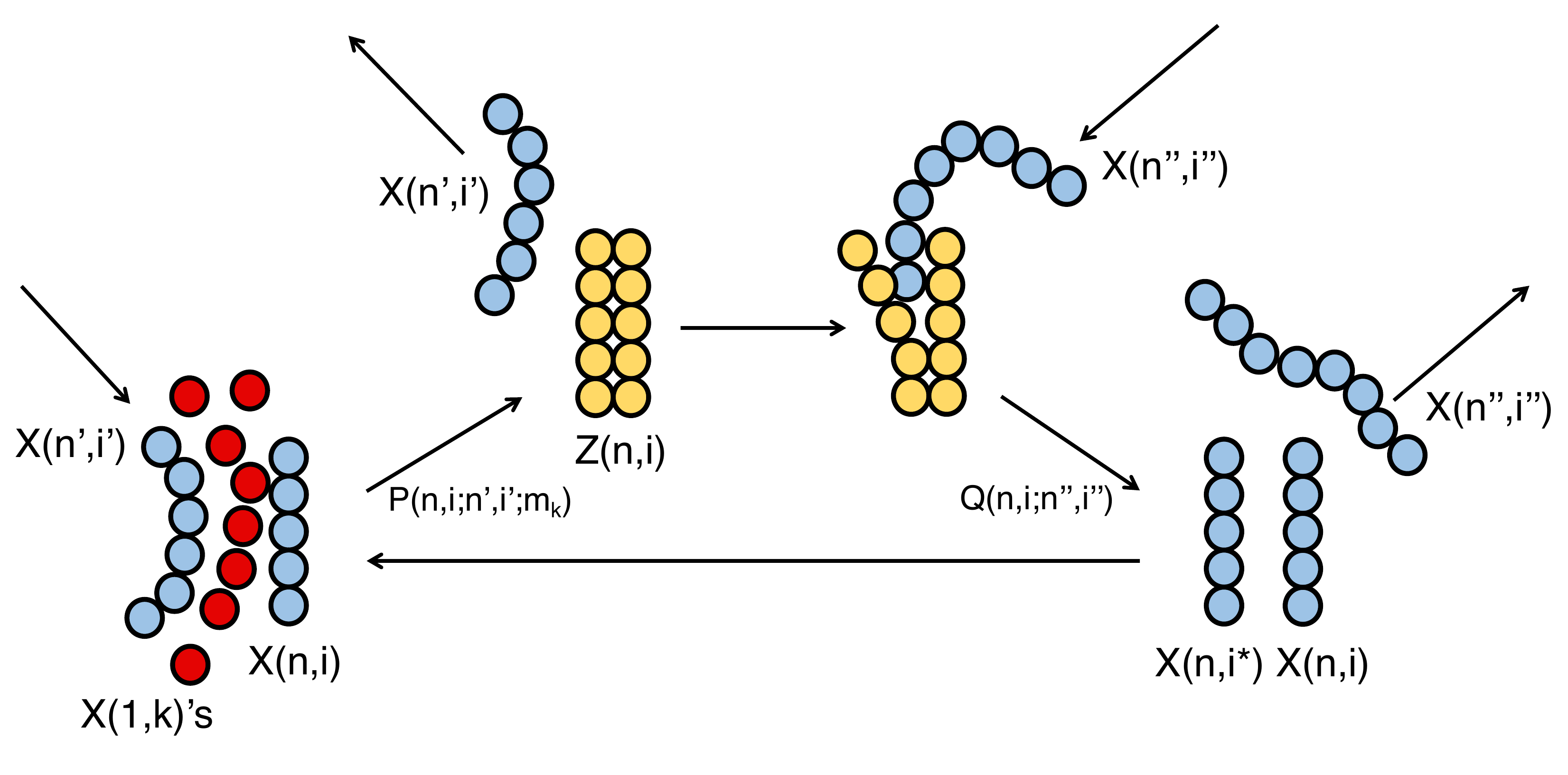}
  \caption{
    Diagram of a pn-molecule and double strand with other pn-molecules and mn-molecules in the beginning pre-RNA world just after the transition from material world. The pn-molecules, mn-molecules and double strand are shown by the blue, red and yellow colors, respectively. The pn-molecule $X(n,i)$ and $X(n,i^*)$ under consideration are shown by the vertical molecules and the other interacting molecules $X(n',i')$ and $X(n'',i'')$ are shown by the slanted molecules. The arrows indicate the directions of reactions and the flows of interacting pn-molecules.
  }
\label{fig:model2}
\end{center}
\end{figure*}

On the time axis of material world with increasing density of pn-nucleotide molecules, the chance increased of each pn-nucleotide molecule interacting with other pn-molecules which might have contributed to forming a double strand $Z(n,i)$ as shown in Fig.~\ref{fig:model2}.  Although the double strand $Z$ is known rather stable in the laboratory experiment, it might have been separated into a pn-nucleotide molecule $X(n,i)$ and its compliment molecule $X(n,i^*)$ spontaneously and/or by the help of surrounding pn-molecules under some off-laboratory condition \cite{szostak12}.

It is noted that the model which represents the scenario described here and in the previous sections should be limited by the following requirements:

(i) To discuss a transition from the material world to the pre RNA-world, use of special function of molecules should not be included: Ligase and other ribozymes, which have been thought to function in the RNA world, would not exist before this transition. Although we use terminology `mutually catalytic', it is equivalent to `mutually interactive'. It does not mean any molecules with special functions.

(ii) The activation of self-replicator is indispensable for the change from the material world. There is no robust self-replicator in any material worlds. Therefore, the birth of self-replicator can be interpreted as the origin of the RNA life. Sharp growth of high fidelity informative pn-molecules is achieved only by self-replicators.

(iii) The transition must occur at a specific point on the time axis of material world. This condition is not satisfied by linear dynamics of spontaneous reaction, because the molecular density would increase or decrease exponentially regardless of pn-molecule density. Also, nonlinear dynamics is essential to avoid the poor information quality of linear dynamics at the separation of double strands \cite{eigen71}. Spontaneous separation of double strands would occur in parallel with their natural decay, and the resultant molecules will suffer statistical error in the information.

(iv) The higher degree of nonlinearity of the differential equation is utilized, the higher density of pn-molecules world at the critical transition is generally required. Ligation would require third order nonlinearity.  A second order differential equation, which corresponds to a dynamical system of interacting two molecules, is suited for representing the first transition from material world.

We present hereafter a dynamical onset model of mutually catalytic self-replication which satisfies these four requirements. As shown in Fig.~\ref{fig:model2}, the pn-molecule $X(n,i)$ and the corresponding double strand $Z(n,i)$ under consideration are surrounded by environmental pn-molecule molecules such as $X(n',i'), X(n'',i''), ...,$ and mn-molecules $X(1,k)$'s in the material world. It was assumed in this model that there is one pn-molecule $X(n',i')$ which interacts most strongly with $X(n,i)$ with reaction constant $P$ to form $Z(n,i)$ and one pn-molecule $X(n'',i'')$ which interacts most strongly with $Z(n,i)$ with reaction constant $Q$ to separate it into $X(n,i)$ and $X(n,i^*)$.

The reactions of the pn-molecules $X(n,i)$ and the double strand $Z(n,i)$ are written with chemical reaction constants $P$ and $Q$ as
\begin{align}
  &X(n,i) + X(n',i') + \sum_{k=1}^{4} m_k X(1,k) \nonumber \\
  &~~~~~~~~\xrightarrow{P(n,i;n',i';m_k)} Z(n,i) + X(n',i'),\label{eq:xn}\\
  &Z(n,i) + X(n'',i'') \nonumber \\
  &~~~~~~~~\xrightarrow{Q(n,i;n'',i'')} X(n,i) + X(n,i^*) + X(n'',i''), \label{eq:zn}
\end{align}
where $m_k$ is the number of $k$-th nucleoside in the pn-molecule $X(n,i)$, and therefore, $\sum_{k=1}^4 m_k = n$. In the Eqs.~(\ref{eq:xn}) and (\ref{eq:zn}) the spontaneous reactions for formation and separation of double strand $Z$ are not included by taking the requirement (iii) into account.
     
When the density of monomer $X(1,k)$ is high and saturated at $C_s$, $\prod_{k=1}^{4} X(1,k)^{m_k}$ will stay at $(C_s)^n$ during the onset time, and the dynamics of the reactions will be written as,
\begin{align}
  \frac{dZ(n,i)}{dt} = (C_s)^n P(n,i;n',i') X(n,i) X(n',i').
  \label{eq:dzn2}
\end{align}

\begin{figure}
\begin{center}
  \includegraphics[height=4.7cm,keepaspectratio]{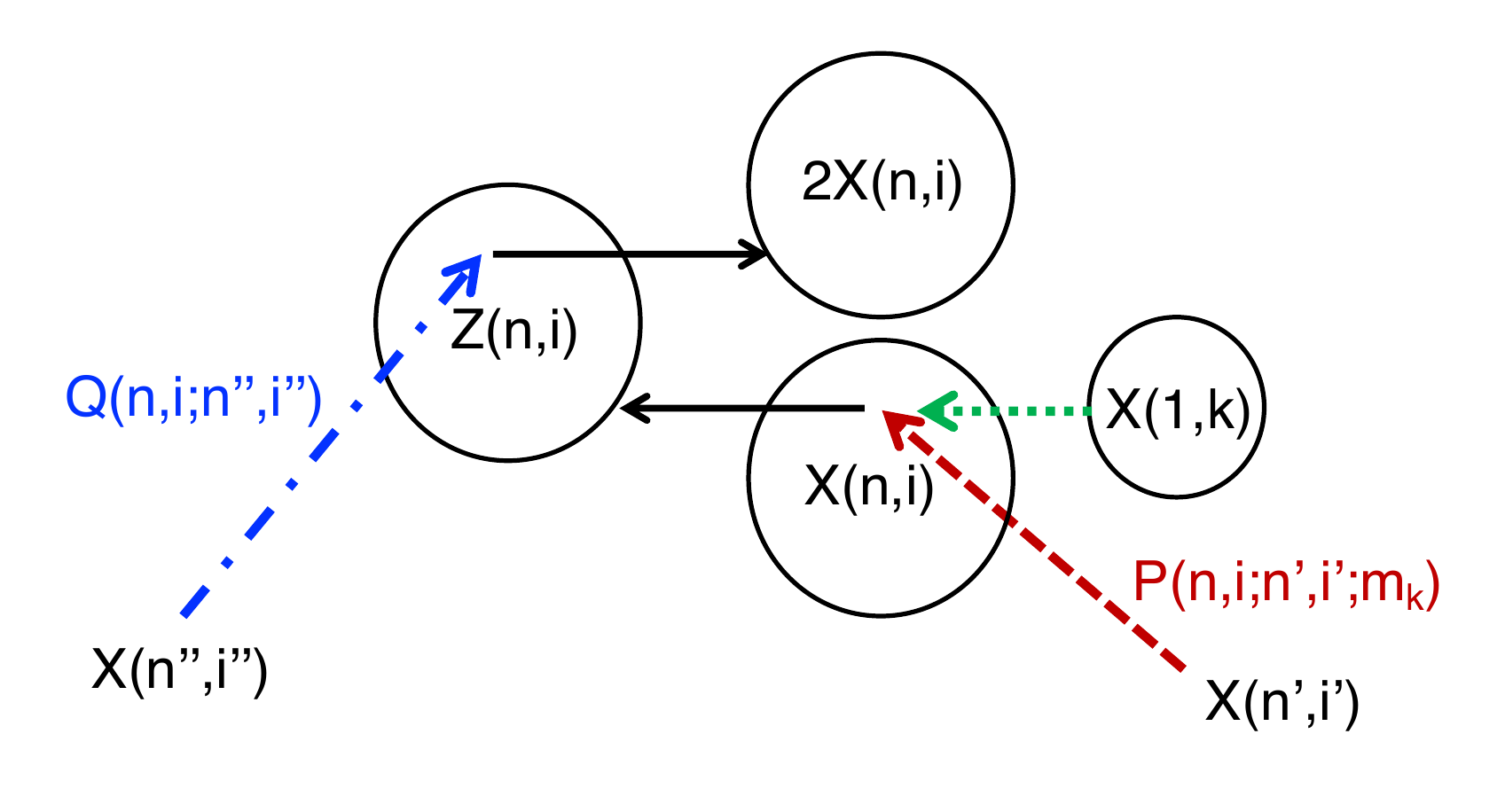}
  \caption{Mutually catalytic self-replication models by double strands without considering complements. The red arrow means catalytic reaction of $X(n',i')$ for copying the template $X(n,i)$ by $X(1,k)$ with a reaction constant $P(n,i;n',i';m_k)$. The blue arrow for separating a double strand $Z(n,i)$ to two single strands $X(n,i)$ with a reaction constant $Q(n,i;n'',i'')$. The black arrows indicate the result of the reactions.
}
\label{fig:single}
\end{center}
\end{figure}

Because $(n,i^*)$ belongs to the set $\{n,i\}$ and vice versa, the density of complement polynucleotide $X(n,i^*)$ may be assumed to be equal to the density $X(n,i)$ according to Eq.~(\ref{eq:zn}), neglecting possible small difference of initial condition. When we discuss a framework in which $X(n,i)$ is meant by $X(n,i)$ plus its complement $X(n,i^*)$, the reaction (\ref{eq:zn}) is rewritten as
\begin{equation}
  Z(n,i) + X(n'',i'') \xrightarrow{Q(n,i;n'',i'')} 2 X(n,i) + X(n'',i''),
\end{equation}
and shown in Fig.~\ref{fig:single}. The corresponding dynamical equation is written as
\begin{equation}
  \frac{dX(n,i)}{dt} = \frac{1}{2} Q(n,i;n'',i'') Z(n,i) X(n'',i'').
  \label{eq:dxn2}
\end{equation}
Inclusion of more than one $X(n',i')$ in Eq.~(\ref{eq:xn}) or more than one $X(n'',i'')$ in Eq.~(\ref{eq:zn}) would make the dynamical equation (\ref{eq:dzn2}) or (\ref{eq:dxn2}) with higher nonlinearity, which is avoided by the requirement (iv).

We use a simplified symbol hereafter,
\begin{equation}
  X(n,i) \to X_\nu, ~~~ Z(n,i) \to Z_\nu,
\end{equation}
where $\nu$ represents both the molecular length $n$ and the information sequence $i$ in Eqs.~(\ref{eq:dzn2}) and (\ref{eq:dxn2}).
Similarly,
\begin{equation}
  (C_s)^n P(n,i;n',i')
  \to p(\nu,\nu'), ~~~ \frac{1}{2}Q(n,i;n'',i'') \to q(\nu,\nu'').
  \label{eq:original2}
\end{equation}
Explicitly writing time $t$ for the variables, we obtain a set of nonlinear differential equations for $Z_\nu(t)$ and $X_\nu(t)$ as
\begin{align}
  \frac{dZ_\nu(t)}{dt} = p(\nu,\nu') X_\nu(t) X_{\nu'}(t) -\frac{Z_\nu(t)}{\tau_{z,\nu}}, \label{eq:dzt} \\
  \frac{dX_\nu(t)}{dt} = q(\nu,\nu'') Z_\nu(t) X_{\nu''}(t) -\frac{X_\nu(t)}{\tau_{x,\nu}}, \label{eq:dxt}
\end{align}
where we have added natural decay terms of the variables.

One notices that when the concentration of mononucleotides is not saturated, $(C_s)^n$ in the dynamics of Eq.~(\ref{eq:dzn2}) is replaced by $C^n$, and thereby Eq.~(\ref{eq:dzt}) is written, by using the same $p(\nu,\nu')$ in Eq.~(\ref{eq:original2}), as
\begin{equation}
  \frac{dZ_\nu(t)}{dt} = \left(\frac{C}{C_s}\right)^n p(\nu,\nu') X_\nu(t) X_{\nu'}(t) - \frac{Z_\nu(t)}{\tau_{z,\nu}}.
  \label{eq:dzt_r}
\end{equation}
The analysis of Eqs.~(\ref{eq:dzt}) and (\ref{eq:dxt}) for saturated mononucleotides density will be described in the following sections, and the dynamics of Eq.~(\ref{eq:dxt}) and (\ref{eq:dzt_r}) will be separately be used for the discussion of the length of self-replicated polynucleotide molecules in Section VI.

\begin{figure*}
  \begin{center}
    \includegraphics[height=7.0cm,keepaspectratio]{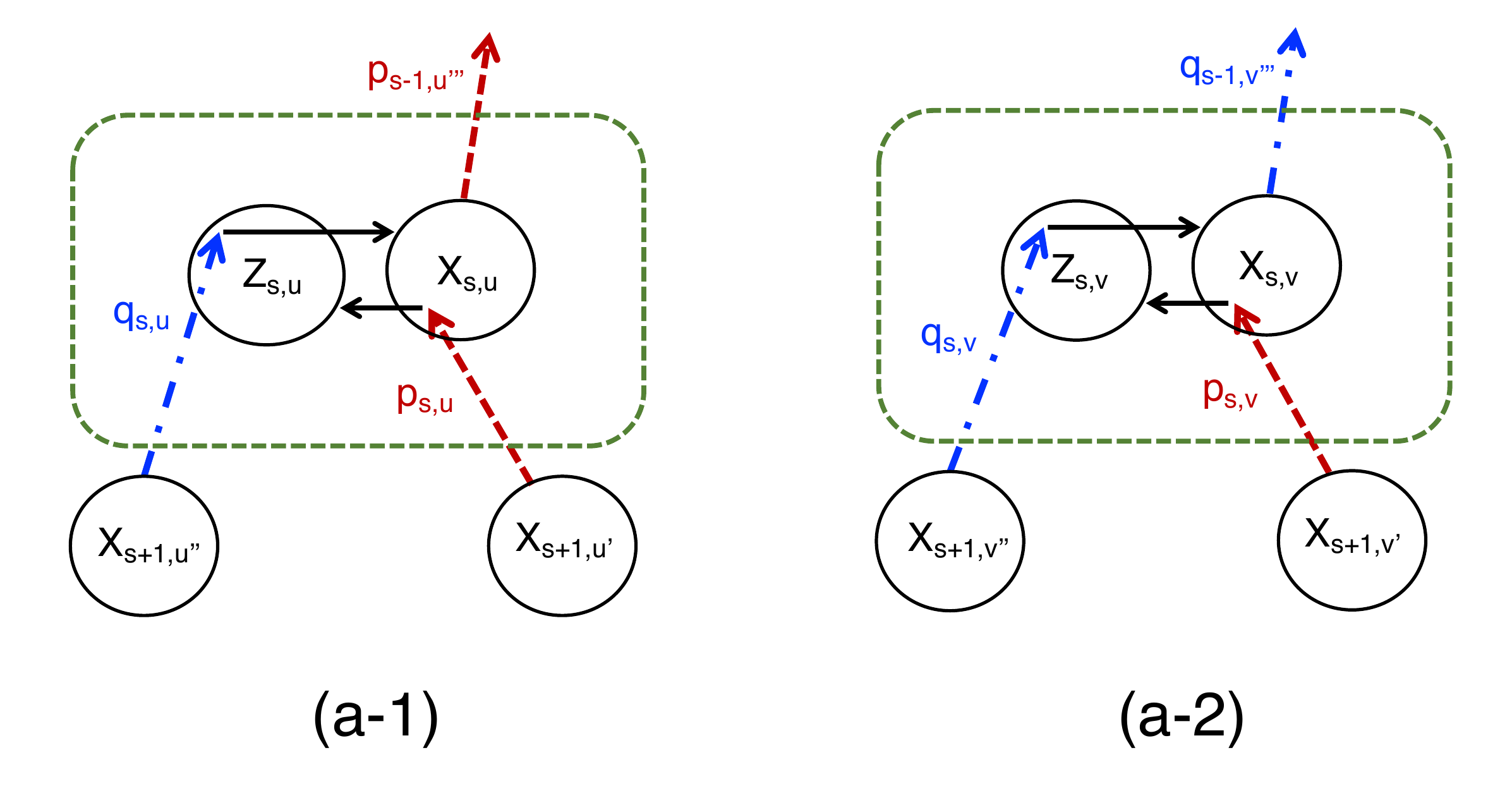}
   \end{center}  
\end{figure*}
\begin{figure*}
   \begin{center}
     \includegraphics[height=7.0cm,keepaspectratio]{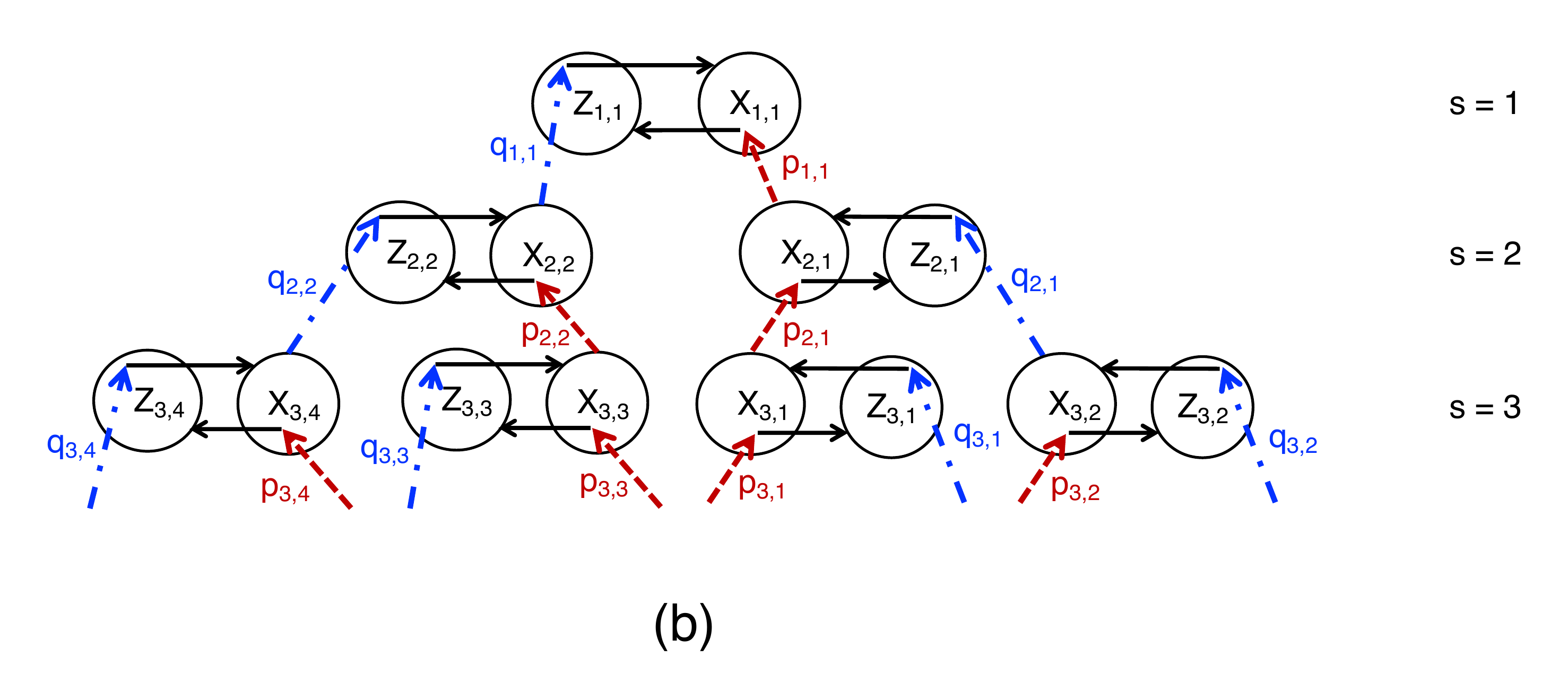}
   \end{center}
   \caption{Open ended self-replicator networks. Two kinds of self-replicator units shown in panels (a-1) and (a-2) are necessary for constituting an open ended network such as shown in (b). Panel (a-1): A p-p type of self-replicator unit. It receives catalytic interactions $p_{s,u}$ (the red arrow) and $q_{s,u}$ (the blue arrow) from other units $(s+1, u')$ and $(s+1, u'')$. It gives a catalytic interaction $p_{s-1,u'''}$ to the other unit $(s-1, u''')$. Panel (a-2): A p-q type of self-replicator unit. It receives catalytic interactions $p_{s,v}$ (the red arrow) and $q_{s,v}$ (the blue arrow) from other units $(s+1, v')$ and $(s+1, v'')$. It gives a catalytic interaction $q_{s-1,v'''}$ to the other unit $(s-1, v''')$. Panel (b): An example of an open ended network constructed by the p-p type and p-q type of self-replicator units. Only a part of network to $s=3$ layer is shown.
}
   \label{fig:open}
\end{figure*}

Even with this formalism the networks are still divided into two cases. The first case is where one kind of pn-molecule $X_\nu$ can have only one catalytic function either replication or separation. The second case is where pn-molecules can have both catalytic replication function on a pn-molecule and a catalytic separation function on another molecule simultaneously. In the first case, the network is described as an open one, while in the second case, the network can be described by a closed network.

To conclude this section, it is worth mentioning that the present model satisfies the requirements (i) to (iv), and therefore, is necessary and sufficient to discuss the onset problem of pre-RNA world.

\begin{figure*}
\begin{center}
  \includegraphics[height=7.0cm,keepaspectratio]{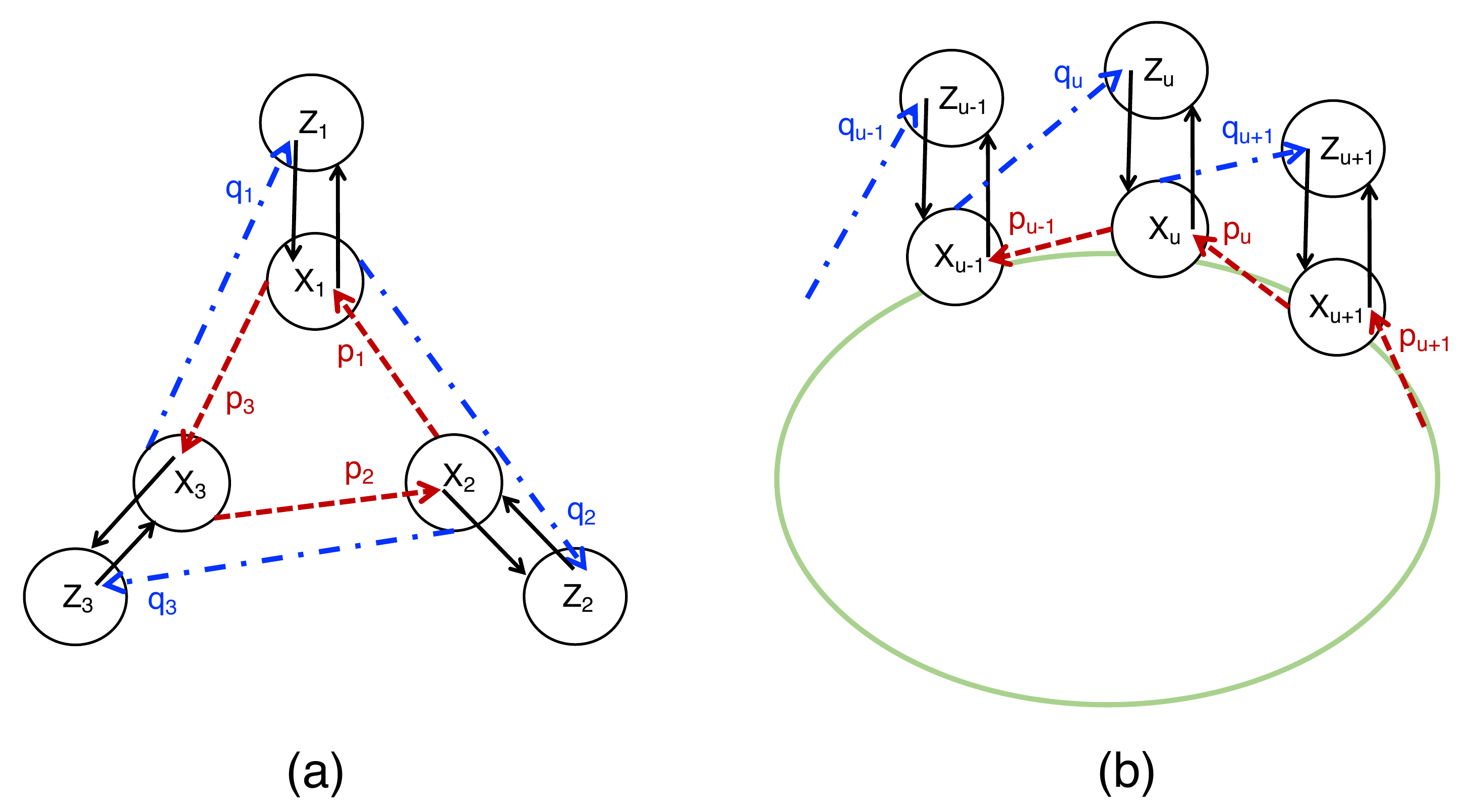}
  \caption{A Closed loop self-replication of networks. Panel (a): A minimum closed network of three kinds of self-replicator units, each of which has two kinds of catalytic interactions, for doubling and for separating shown by a red arrow and by a blue line, respectively. Panel (b): A part of a closed network of mutually catalyzing self–replicators. The network is an extension of Panel (a) to a network of $N$ self-replicator units.
}
\label{fig:closed}
\end{center}
\end{figure*}

\section{III Networks of mutually catalytic pn-molecules}
\subsection{A. Open ended self-replicator networks}

A network of open ended self-replicators has a layer structure composed of two kinds of self-replicator units, p-p type and p-q type, as shown in panels (a-1) and (a-2) of Fig.~\ref{fig:open}, respectively. The p-p type is composed of $X_{s,u}$ and $Z_{s,u}$, and the p-q type of $X_{s,v}$ and $Z_{s,v}$, where $s$ represents the layer number and $u,v$ represent the address of the unit in the layer $s$. $X_{s,u}$ of a p-p type unit receives a catalytic interaction $p_{s,u}$ (the red arrow) from $X_{s+1,u'}$ to form a double strand $Z_{s,u}$ by copying itself, and $Z_{s,u}$ receives a catalytic interaction $q_{s,u}$ (the blue arrow) from $X_{s+1,u''}$ to decompose into two $X_{s,u}{}'$s. Only difference between the two types lies in the catalytic interaction acting an unit in $s-1$ th layer. $X_{s,u}$ of a p-p type unit gives a catalytic interaction $p_{s-1,u'''}$ (the red arrow) to $X_{s-1,u'''}$, while $X_{s,v}$ of a p-q type unit gives a catalytic interaction $q_{s-1,v'''}$ (the blue arrow) to $Z_{s-1,v'''}$. Panel (b) shows an example of an open ended network constructed by the p-p type and p-q type of self-replicator units. The addresses in s-th layer in (b) panel of Fig.~\ref{fig:open} are chosen arbitrarily with a condition $1 \leq u' \neq u'' \neq v' \neq v'' \leq 2^s$.

For a p-p replicator unit in $s$-th step group, the dynamics is
\begin{align}
  &\frac{dZ_{s,u}(t)}{dt} = p_{s,u}X_{s,u}(t)X_{s+1,u'}(t)-\frac{Z_{s,u}(t)}{\tau_z}, \label{eq:pp1}\\
  &\frac{dX_{s,u}(t)}{dt} = q_{s,u}Z_{s,u}(t)X_{s+1,u''}(t)-\frac{X_{s,u}(t)}{\tau_x}.\label{eq:pp2}
\end{align}
For a p-q replicator unit in $s$-th step group, the dynamics is
\begin{align}
  &\frac{dZ_{s,v}(t)}{dt} = p_{s,v}X_{s,v}(t)X_{s+1,v'}(t)-\frac{Z_{s,v}(t)}{\tau_z}, \\
  &\frac{dX_{s,v}(t)}{dt} = q_{s,v}Z_{s,v}(t)X_{s+1,v''}(t)-\frac{X_{s,v}(t)}{\tau_x}.
\end{align}
Although $\tau_x$ may depend in principle on $s, u$, and $\tau_z$ on $s, v$, all these dependences are not considered in the present paper. The analyses and an example of simulation of the open ended network shown in Fig.~\ref{fig:open} (b) is presented in Sec.~IV.

\subsection{B. Closed loop self-replicator networks}

For closed networks, we first consider a simplest network composed of three self-replicator units $[X, Z]$. In each of three units, $X$ has both a catalytic replication function on another pn-molecule and a catalytic separation function on another different molecule simultaneously. The network is shown in panel (a) of Fig.~\ref{fig:closed}, with possibilities of different interaction strengths, $p{}'$s and $q{}'$s.

For networks composed of $N$ self-replicator units $(N>3)$, one can imagine variety of complex networks, but we limit ourselves in this paper only to the simple type shown in panel (b) of Fig.~\ref{fig:closed}, which is an extension of the type shown in Fig.~\ref{fig:closed} (a). This model is based on the assumption that a pn-molecule interacts catalytically with only one of the other molecules of strongest interaction.

The subscript in this case can be simplified without losing generality. A kind of pn-molecule and its double strand can be written as $X_u$ and $Z_u$, where $u$ is the address in the ring. $X_u$ in the ring is assumed to have a catalytic interaction to produce a double strand $Z_u = X_u X_u$ by a neighboring $X_{u+1}$, and the double strand $Z_u$ is simultaneously catalytically reacted by $X_{u-1}$. This occurs for all $u$-th elements of $X_u$ and $Z_u$ from $u=1$ to $N$.

The dynamics is written as
\begin{align}
  &\frac{dZ_u(t)}{dt} = p_u X_u(t) X_{u+1}(t) - \frac{Z_u(t)}{\tau_z},
  \label{eq:closed1}\\
  &\frac{dX_u(t)}{dt} = q_u Z_u(t) X_{u-1}(t) - \frac{X_u(t)}{\tau_x},
  \label{eq:closed2}  
\end{align}
where the quantities with indices $u=0$ and $u=N+1$ are equivalent to those with indices $u=N$ and $u=1$, respectively.
The anlyses of Eqs.~(\ref{eq:closed1}) and (\ref{eq:closed2}) will be shown in Sec.~V.

\section{IV. Analysis and numerical simulation of open ended networks}

\subsection{A. Critical boundary condition analysis for onset of self-replication}

It may be reasonable to imagine an open ended network shown in Fig.~\ref{fig:open} (a) may be spread in a real space, and may meet a spatial boundary at $s=b$. We analyze the boundary value dependence of the activity of an open network which ends at some step $b$. As a simplest example, we assume that all the values in step $s=3$ as fixed boundary values,
\begin{equation}
X_{3,u} = X_{b,u}~~~(u=1,2,3,4).
\end{equation}  
Then Eqs. (\ref{eq:pp1}) and (\ref{eq:pp2}) for $s=2$ are linear differential equations and we can eliminate $Z_{2,1}$ and obtain,
\begin{align}
  &\frac{d^2 X_{2,1}}{dt^2} + (\tau_z^{-1} + \tau_x^{-1})\frac{dX_{2,1}}{dt} \nonumber\\
  &~~~~+ \tau_z^{-1} \tau_x^{-1} (1-p_{2,1}q_{2,1}\tau_z \tau_x X_{b,1}X_{b,2})X_{2,1} = 0.
\end{align}
By setting $X_{2,1} \propto \exp(\lambda t)$, we obtain
\begin{equation}
  \lambda = \frac{1}{2}\left[\sqrt{(\tau_z^{-1}-\tau_x^{-1})^2+4p_{2,1}q_{2,1}X_{b,1}X_{b,2}}-(\tau_z^{-1}+\tau_x^{-1})\right],
\end{equation}
which has a positive solution when
\begin{equation}
  X_{b,1}X_{b,2} > (p_{2,1}q_{2,1}\tau_z\tau_x)^{-1} \equiv (X_{b,1}X_{b,2})_c.
  \label{eq:open_c}
\end{equation}
Here $(X_{b,1}X_{b,2})_c$ is the critical value of $X_{b,1}X_{b,2}$ for which $Z_{2,1}(t)$ and $X_{2,1}(t)$ take stationary values of the dynamics given by Eqs.~(\ref{eq:pp1}) and (\ref{eq:pp2}). When Eq.~(\ref{eq:open_c}) is satisfied, the self-replicator $[X_{2,1}, Z_{2,1}]$ shown in Fig.~\ref{fig:open} (b) starts operating and the values of $X_{2,1}$ and $Z_{2,1}$ increase exponentially.

Likewise when $X_{b,3}$ and $X_{b,4}$ satisfy a condition similar with Eq.~(\ref{eq:open_c}), the self-replicator $[X_{2,2}, Z_{2,2}]$ of the Fig.~\ref{fig:open} (b) will increase. Then after some time when $X_{2,1} X_{2,2}$ satisfy the condition similar to Eq.~(\ref{eq:open_c}), the next self-replicator of $s=1$ step $[X_{1,1}, Z_{1,1}]$ in Fig.~\ref{fig:open} (b) will start increasing like a chain reaction. More generally, in case when the boundary of the open network is given at $b$-th stage from the top, all the self-replication dynamics of $[X_{s,u}, Z_{s,u}]$ for $1 \leq s < b$ up to $[X_{1,1}, Z_{1,1}]$ can be excited, when all the members $X_{b,u}$ for odd $u'$s with $1 \leq u < 2^{b-1}$ satisfy
\begin{equation}
  X_{b,u}X_{b,u+1} > (X_{b,u}X_{b,u+1})_c,
  \label{eq:open_bc}
\end{equation}
where $(X_{b,u}X_{b,u+1})_c$ is the critical value of $X_{b,u}X_{b,u+1}$ which makes $[X_{b-1,u}(t) , Z_{b-1,u}(t)]$ stationary, and is equal to $(p_{b-1,u}q_{b-1,u}\tau_z\tau_x)^{-1}$.

\subsection{B. Numerical simulation with given boundary conditions}

\begin{figure}
  \begin{center}
    \includegraphics[scale=0.6]{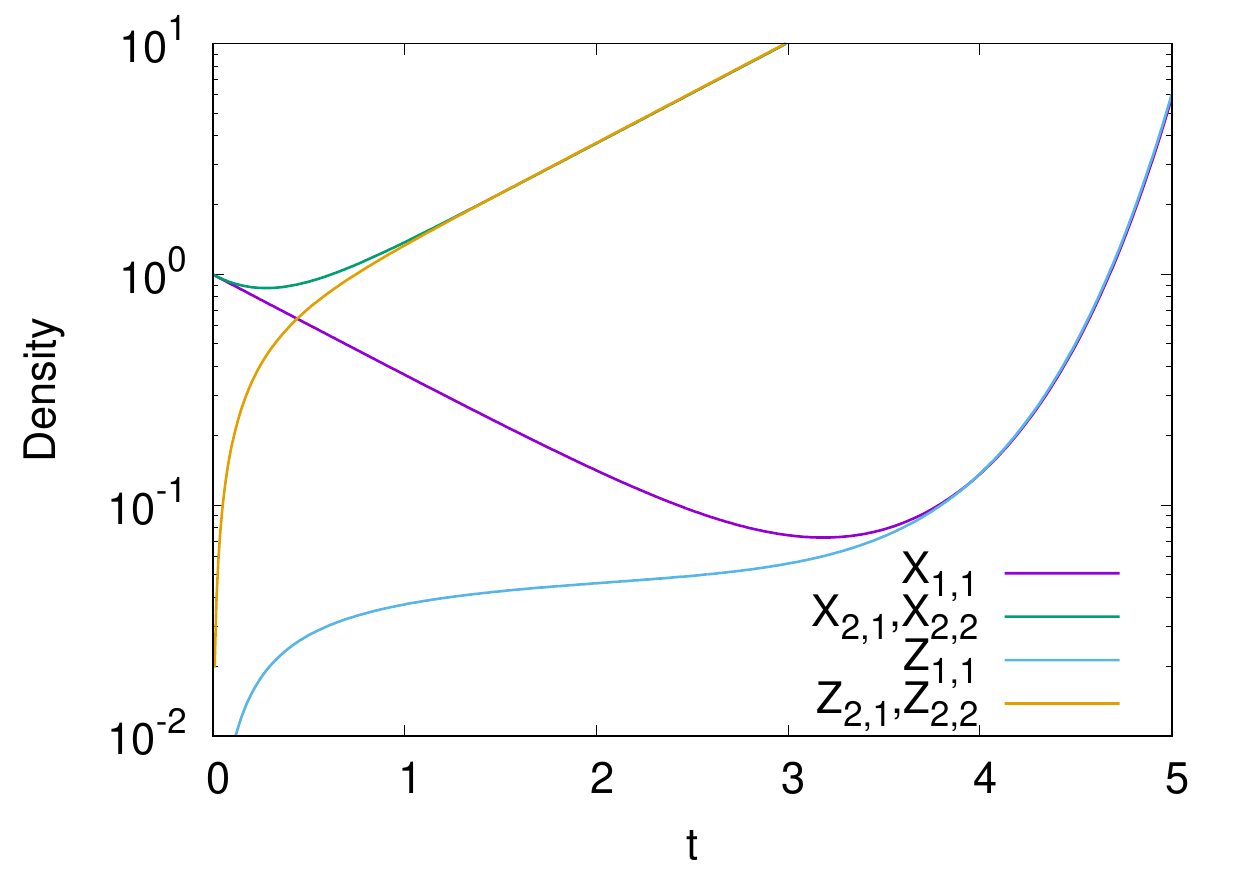}
  \end{center}
  \caption{Temporal variation of the top two layers of an open ended network. The boundary values at $s=3$ are set to be $X_{b,1}=X_{b,2}=X_{b,3}=X_{b,4}=20$, which satisfies Eq.~(\ref{eq:open_bc}). The initial values are $X_{1,1}(0)=X_{2,1}(0)=X_{2,2}(0)=1, Z_{1,1}(0)=Z_{2,1}(0)=Z_{2,2}(0)=0$.}
  \label{fig:open_s}
\end{figure}

\begin{figure}
  \begin{center}
    \includegraphics[scale=0.6]{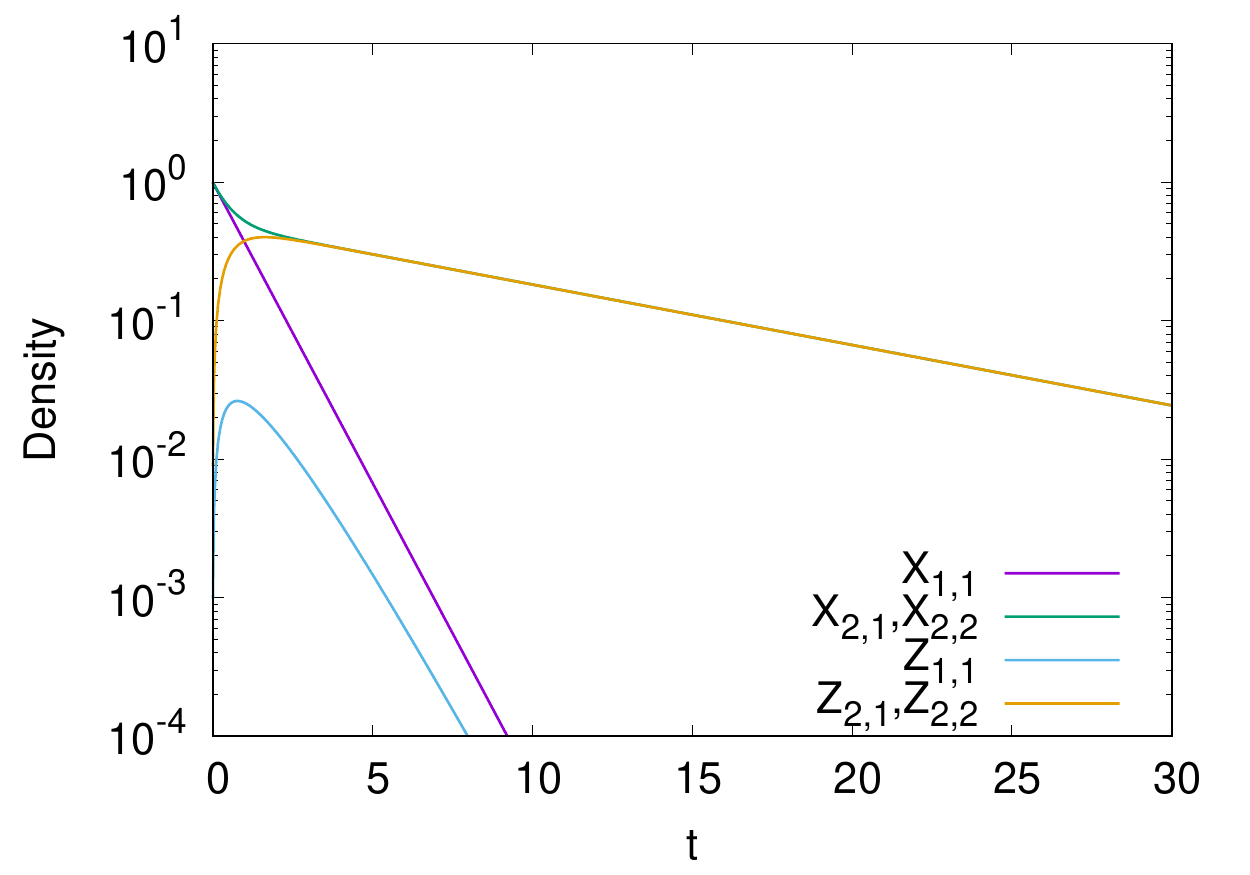}
  \end{center}
  \caption{Temporal variation of the top two layers of an open ended network. The boundary values at $s=3$ are set to be $X_{b,1}=X_{b,2}=X_{b,3}=X_{b,4}=9$, which does not satisfy Eq.~(\ref{eq:open_bc}). The initial values are $X_{1,1}(0)=X_{2,1}(0)=X_{2,2}(0)=1, Z_{1,1}(0)=Z_{2,1}(0)=Z_{2,2}(0)=0$.}
  \label{fig:open_s2}
\end{figure}

We carried out numerical simulations of the open ended mutually catalytic network of self-replication, assuming the members in $s=3$ as fixed boundaries $X_{b,1}, X_{b,2}, X_{b,3},$ and $X_{b,4}$. The dynamics of the members in $s=1$ and $s=2$ layers in Fig.~\ref{fig:open} (b) is
\begin{align}
  &\frac{dZ_{1,1}(t)}{dt} = p_{1,1}X_{1,1}(t)X_{2,1}(t) - \frac{Z_{1,1}(t)}{\tau_z},\nonumber\\
  &\frac{dX_{1,1}(t)}{dt} = q_{1,1}Z_{1,1}(t)X_{2,2}(t) - \frac{X_{1,1}(t)}{\tau_x},\nonumber\\
  &\frac{dZ_{2,1}(t)}{dt} = p_{2,1}X_{b,1}X_{2,1}(t) - \frac{Z_{2,1}(t)}{\tau_z},\nonumber\\
  &\frac{dX_{2,1}(t)}{dt} = q_{2,1}X_{b,2}Z_{2,1}(t) - \frac{X_{2,1}(t)}{\tau_x},\nonumber\\
  &\frac{dZ_{2,2}(t)}{dt} = p_{2,2}X_{b,3}X_{2,2}(t) - \frac{Z_{2,2}(t)}{\tau_z},\nonumber\\
  &\frac{dX_{2,2}(t)}{dt} = q_{2,2}X_{b,4}Z_{2,2}(t) - \frac{X_{2,2}(t)}{\tau_x}.\nonumber
\end{align}
We set all the values of $p_{s,u}$ and $q_{s,u}$ as $0.1$, and $\tau_z=\tau_x=1$, which give $(X_{2,1}X_{2,2})_c = (X_{b,1}X_{b,2})_c = (X_{b,3}X_{b,4})_c = 100$. As examples, we examined two cases, $X_{b,1}=X_{b,2}=X_{b,3}=X_{b,4}=20$ and $X_{b,1}=X_{b,2}=X_{b,3}=X_{b,4}=9$. The former case satisfies Eq.~(\ref{eq:open_bc}), while the latter case does not. The simulation results are shown in Figs.~\ref{fig:open_s} and \ref{fig:open_s2}. We set the initial values as $X_{1,1}(0)=X_{2,1}(0)=X_{2,2}(0)=1$ and $Z_{1,1}(0)=Z_{2,1}(0)=Z_{2,2}(0)=0$ for both of the cases.

As expected in Sec.~IV A., $X_{2,1}$ and $X_{2,2}$ grow, and $X_{1,1}(t)$ begins to be stationary at $t \sim 3$ in Fig.~\ref{fig:open_s}, when $X_{2,1}(t)=X_{2,2}(t) \sim 10$ which satisfies “critical boundary condition” Eq.~(\ref{eq:open_bc}) for the dynamics of $X_{1,1}$ and $Z_{1,1}$. This is an example of chain reaction-like behavior of the cascade type open self-replication network. On the other hand, $X_{1,1}$ and $Z_{1,1}$ have no chance to be stationary as seen in Fig.~\ref{fig:open_s2}, because the set of the boundary values of the boundary layer values do not satisfy the condition Eq.~(\ref{eq:open_bc}). Comparison of Figs.~\ref{fig:open_s} and \ref{fig:open_s2} demonstrates that the criteria of the growth of connected pn-molecules given by Eq.~(\ref{eq:open_bc}) is justified.

\section{V. Analysis and numerical simulation of closed loop networks}
\subsection{A. Stationery value analysis of three self-replication units}

The mutually catalytic interaction scheme of a closed network are shown for three self-replication units case in Fig.~\ref{fig:closed} (a). We look for the stationary point of this dynamics for $X_u$ and $Z_u$ with $u=1,2,3$.

The dynamics of the model is given by Eqs.~(\ref{eq:closed1}) and (\ref{eq:closed2}), where the quantities with indices $u=0$ and $u=4$ are equivalent to those with indices $u=3$ and $u=1$, respectively. The stationary point of the system is obtained as,
\begin{align}
  &X_u^* = p_u q_u \left(\tau_z\tau_x \prod_{\ell=1}^3 p_\ell q_\ell \right)^{-1/2}, \\
  &Z_u^* = p_u /(\tau_x p_{u-1} q_{u-1}).
\end{align}
As a simple example when $q_u = q, \tau_z =\tau_x =\tau$, the stationary point can be written more concretely as,
\begin{align}
  &X_u^* = [p_u/(\tau^2 q p_{u+1} p_{u+2})]^{1/2}, \label{eq:xi*}\\
  &Z_u^* = p_u/(\tau q p_{u-1}). \label{eq:zi*}
\end{align}

\begin{table*}
  \caption{\label{tab:closed_s}Initial conditions $X_u(0)$ and $Z_u(0)$ and simulation results (grow or decay) in Fig.~\ref{fig:closed_s}. $X_g(0)$ and $Z_g(0)$ are the geometric mean of the initial values, and $X_u^*$ and $Z_u^*$ are the stationary values. No.~8 in the Table is not shown in Fig.~\ref{fig:closed_s}, due to the largeness of the domain of the flowline.}
  \begin{ruledtabular}
    \begin{tabular}{l|llllll|ll|ll|l}
      No. & $X_1(0)$ & $X_2(0)$ & $X_3(0)$ & $Z_1(0)$ & $Z_2(0)$ & $Z_3(0)$ & $X_g(0)$ & $Z_g(0)$ & \# of $X_u(0)>X_u^*$ & \# of $Z_u(0)>Z_u^*$ & Grow or Decay\\
      \hline
      1 & 300 & 100 & 30 & 0 & 0 & 0 & 97 & 0 & 3 & 0 & G\\
      2 & 210 & 52  & 13 & 0 & 0 & 0 & 52 & 0 & 3 & 0 & D\\
      3 & 150 & 100 & 30 & 0 & 0 & 0 & 76.6 & 0 & 2 & 0 & D\\
      4 & 160 & 100 & 30 & 0 & 0 & 0 & 78.3 & 0 & 2 & 0 & G\\
      5 & 0   & 100 & 30 & 300 & 0 & 0 & 0 & 0 & 2 & 0 & D\\  
      6 & 0   & 100 & 30 & 500 & 0 & 0 & 0 & 0 & 2 & 1 & G\\
      7 & 130 & 45 & 100 & 0 & 0 & 0 & 83.7 & 0 & 1 & 0 & G\\
      8 & 1000 & 45 & 10 & 0 & 0 & 0 & 76.6 & 0 & 1 & 0 & G\\
    \end{tabular}
  \end{ruledtabular}
\end{table*}

\subsection{B. Growth factor of a closed network with $N$ self-replicator units}

The dynamics of a closed network with $N$ self-replicator units are represented by Eqs.~(\ref{eq:closed1}) and (\ref{eq:closed2}), and the stationary point $(X_u^*, Z_u^*)$ is given by coupled equations,
\begin{align}
  p_u X_u^* X_{u+1}^* - Z_u^*/\tau_z = 0, \\
  q_u Z_u^* X_{u-1}^* - X_u^*/\tau_x = 0,
\end{align}
where $X_{N+1}^* = X_1^*$, $Z_{N+1}^* = Z_1^*$, $X_0^* = X_N^*$, and $Z_0^* = Z_N^*$.

Although one cannot obtain the value of $(X_u^*, Z_u^*)$ analytically in general, the geometric mean value $X_g^*$ and $Z_g^*$ are available as,
\begin{align}
  &(X_g^*)^N \equiv \prod_{u=1}^{N} X_u^* = \left(\tau_z^N \tau_x^N \prod_{u=1}^{N}p_u q_u \right)^{-1/2}, \\
  &(Z_g^*)^N \equiv \prod_{u=1}^{N} Z_u^* = \left(\tau_x^N \prod_{u=1}^{N}q_u \right)^{-1}.
\end{align}

Linearized equations of (\ref{eq:closed1}) and (\ref{eq:closed2}) are obtained by setting $X_u(t) = X_u^* [1+\delta X_u(t)]$ and $Z_u(t) = Z_u^* [1+ \delta Z_u(t)]$ as
\begin{align}
  &\tau_z \frac{d \delta Z_u}{dt} = \delta X_{u+1} + \delta X_u - \delta Z_u, \label{eq:linearized1}\\
  &\tau_x \frac{d \delta X_u}{dt} = \delta X_{u-1} + \delta Z_u - \delta X_u,~~~(u=1,2,\dots,N) \label{eq:linearized2}
\end{align}
independently of the values of $p_u$ and $q_u$. Summing these equations from $u=1$ to $N$, and putting $\sum_{u=1}^N \delta X_u = A, \sum_{u=1}^N \delta Z_u = B$, we obtain
\begin{align}
  &\tau_z \frac{dB}{dt} = 2A- B, \\
  &\tau_x \frac{dA}{dt} = B.
\end{align}
By putting $A, B \propto \exp(\lambda t)$, we obtain
\begin{equation}
  \lambda^2 + (1/\tau_z)\lambda - 2/(\tau_z\tau_x) = 0,
\end{equation}
which has one positive solution
\begin{equation}
  \lambda = (1/2\tau_z)[(1+8\tau_z/\tau_x)^{1/2}-1] > 0.
  \label{eq:closed_eigen}
\end{equation}
This implies that the total density of $X$ and $Z$ grow exponentially, independently of the variation of $p_u$ and $q_u$.
It may be interesting to notice that the value of $\lambda$ for $A$ and $B$ is identical to that for the one-kind of self-catalytic self-replicator unit.
Mode selection analysis of the growth pattern of $N$ pn-molecules network is presented in Appendix.

\subsection{C. Numerical simulations of a minimal model}

\begin{figure}
    \begin{center}
      \includegraphics[scale=0.75]{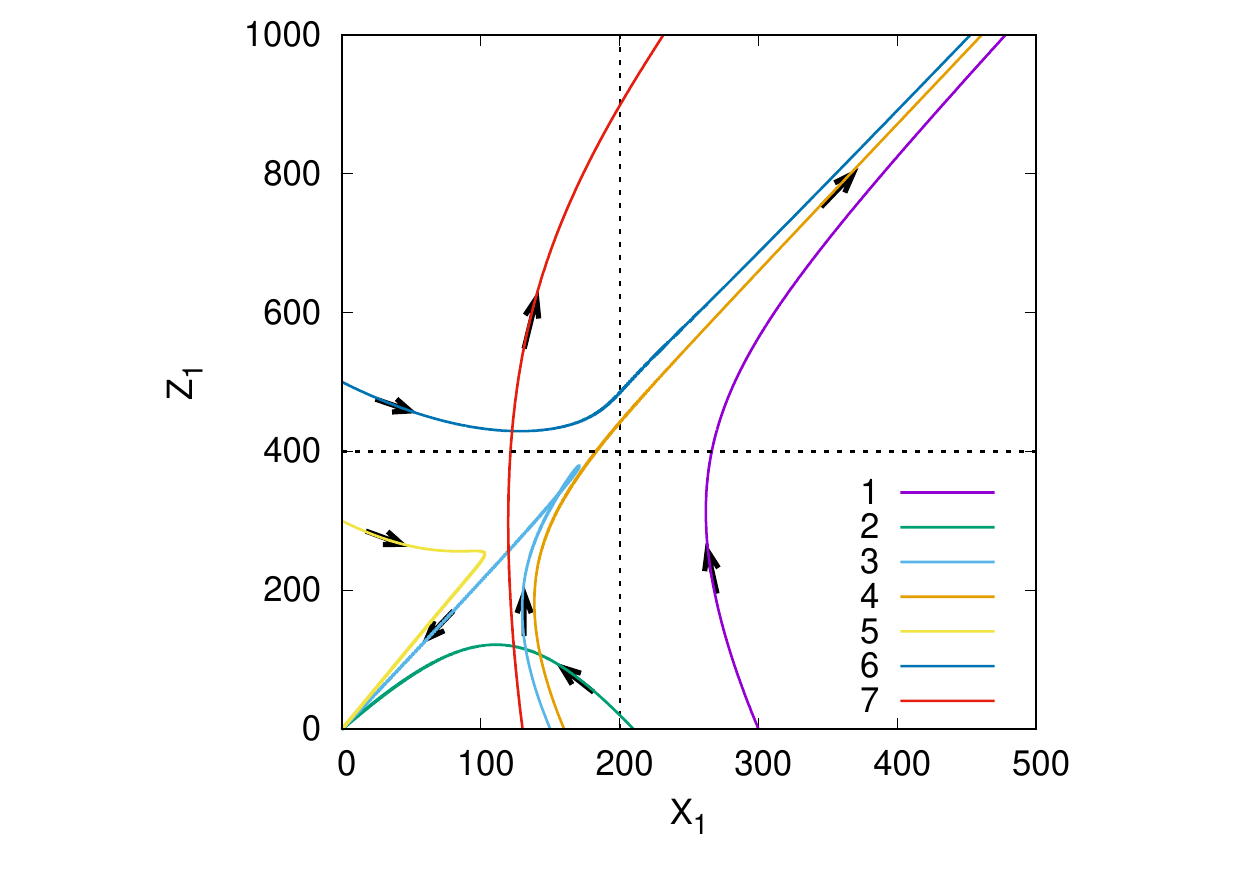}
    \end{center}  
  \caption{Flowlines of the dynamics of the network of three kinds of self-replicator units shown in Fig.~\ref{fig:closed} (a), projected on the plane of $(X_1, Z_1)$. The numbers for the lines correspond to the different initial values $X_u(0)$ and $Z_u(0)$ shown in Table~I. The parameters used in this simulation is $p_1 = 0.16, p_2 = 0.04, p_3 = 0.01, q = 0.01, \tau=1$. The dotted line represents the projected stationary point $(X_1^*, Z_1^*) = (200, 400)$.
  }
  \label{fig:closed_s}
\end{figure}

We show in Fig.~\ref{fig:closed_s} an example of numerical simulation for the dynamics of Eqs.~(\ref{eq:closed1}) and (\ref{eq:closed2}) with various initial conditions shown in Table~I. The parameters used in this simulation is $p_1 = 0.16, p_2 = 0.04, p_3 = 0.01, q = 0.01, \tau=1$, which correspond to the stationary point $(X_1^*, X_2^*, X_3^*, Z_1^*, Z_2^*, Z_3^*) = (200, 50, 12.5, 400, 400, 6.25)$ according to Eqs.~(\ref{eq:xi*}) and (\ref{eq:zi*}). Geometric mean of the stationary position $(X_g^*, Z_g^*) = (50, 100)$.
The numbers for the lines in Fig.~\ref{fig:closed_s} correspond to the different initial values $X_u(0)$ and $Z_u(0)$ shown in Table~I.

The results of simulation for various set of initial values shown in Table~I are summarized as follows.

(1) When the geometric mean $X_g(0)$ of the initial values $X_i(0)'$s exceeds $X_g^* = 50$ over $\sim 50\%$,
$X_u$ and $Z_u$ generally grow (No.~1, 4, and 7), even when some $X_u(0)$ is smaller than $X_u^*$. Otherwise, it decays, even when all $X_u(0)'$s are larger than each $X_u^*$ as in the case of No.~2. Generally speaking, the growth condition is given by,
\begin{equation}
  X_g(0) > r X_g^*, ~~~ r \sim 1.5
  \label{eq:closed_bc}
\end{equation}
This condition is verified even for the case $N=1$ as shown later in Fig.~\ref{fig:closed_s1}.

(2) When $X_g(0) \sim 77$, the behavior is critical. For example, No.~3 decays while No.~8 grows even though these cases have the same $X_g(0)=76.6$. For No.~3, three initial values are larger than three stationery values, while for No.~8, two initial values are smaller than the two stationery values, but one initial value $X_1(0) \gg X_1^*$.

(3) Whether the system grows or decays is sensitive to the initial values of $X_u(0)$, but generally not to the initial values of $Z_u(0)$. However, whether a value of $Z_u(0)$ is greater or smaller than its stationary $Z_u^*$ value creates a difference of growth (No.~6) or decay (No.~5), in case $X_g(0)=0$.
The temporal variations of simulated results show
that the growth period of No.~6 appears much later than that of No.~1.
We shall discuss initial conditions for starting self-replication dynamics in the natural process of the development of pn-molecules in Sec.~VI-B., where we consider that the conditions such as $X_1(0) = 0$ and $Z_1(0) \neq 0$ are not realistic, since $Z_u$ must be formed by matching of monomers with $X_u$ as introduced in Sec.~II-B.

\section{VI. Disucssion}

\subsection{A. Critical conditions for the onset of self-replicator networks}

The results of the critical condition for the onset of self-replication dynamics obtained in Sections IV and V are summarized as follows.

(1) When the boundary of the open network is given at $b$-th step from the top, all the self-replication dynamics of $(X_{s,u}, Z_{s,u})$ for $1 \leq s < b$ up to $(X_{1,1}, Z_{1,1})$ can be excited, only when all the members $X_{b,u}$ with $1 \leq u \leq 2^{s-1}$ in the boundary layer satisfy Eq.~(\ref{eq:open_bc}), or
\begin{equation}
  X_g(b,u) > (p_{b-1} q_{b-1} \tau_z \tau_x)^{-1/2},
  \label{eq:open_cd}
\end{equation}
where $X_g(b,u) \equiv (X_{b,u} X_{b,u+1})^{1/2}$ is the geometric mean of $X_{b,u}$ and $X_{b,u+1}$ for all odd numbers of $u$ at the boundary layer $b$. The right hand side is the critical value which makes $[X_{b-1,u}(t), Z_{b-1,u}(t)]$ stationary.

(2) For a closed network we analyzed, the critical condition for the onset of self-replication was given as Eq.~(\ref{eq:closed_bc}), or
\begin{equation}
  X_g(0) > r X_g^* = r \left[\tau_z \tau_x \left(\prod_{u=1}^N p_u q_u\right)^{1/N}\right]^{-1/2}, ~~~ r\sim 1.5,
  \label{eq:closed_cd}
\end{equation}
Eq.~(\ref{eq:open_cd}) is the conditions necessary for all the boundary pairs of an open network, simultaneously and independently from each other, while  Eq.~(\ref{eq:closed_cd}) is the initial condition for a closed network of $N$ pn-molecules. Although we forcus hereafter on the latter, a similar discussion will be possible for the former, because these two conditions are formally alike except for the factor $r$.

\begin{figure}
    \begin{center}
      \includegraphics[scale=0.75]{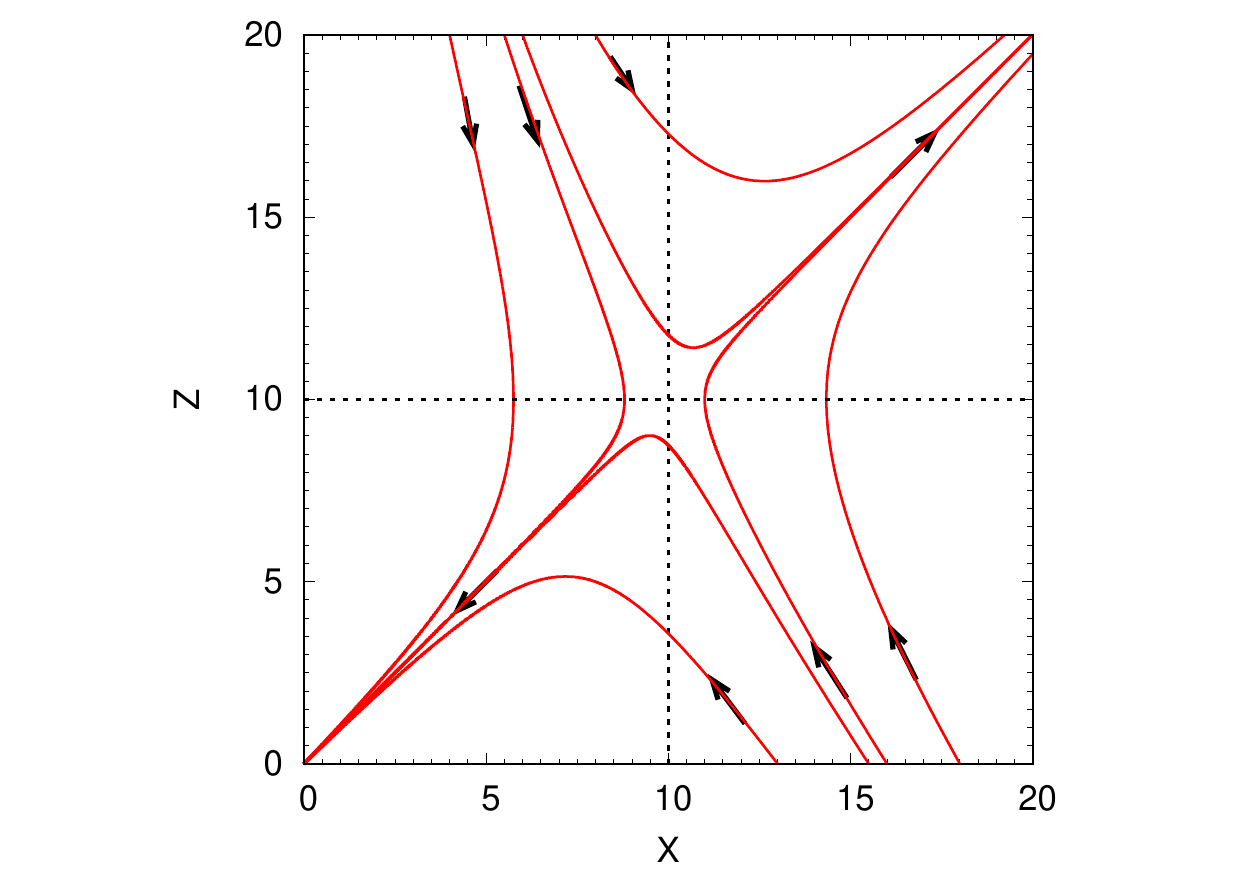}
    \end{center}  
    \caption{Flowlines of $X$ and $Z$ of the $N=1$ model. The parameters are $p=q=0.1, \tau_z=\tau_x=1$. The stationary point is shown by the cross point at $(X^*, Z^*) = (10, 10)$ which is a saddle node of the flow. Critical initial value $X_c(0)$ for growth can be seen from the figure close to $15$, $\sim 50\%$ greater than $X^*$.
  }
  \label{fig:closed_s1}
\end{figure}

To understand the value of $r$ in Eq.~(\ref{eq:closed_cd}), we carried out a simulation of $N=1$ case with $p=q=0.1$ and $\tau_z = \tau_x = 1$. The dynamics is $dZ/dt = pX^2 - Z/\tau_z,$ and $dX/dt = q ZX - X/\tau_x.$
The stationary point in this case is $X^* = Z^*= 10$. The calculated flowlines shown in Fig.~\ref{fig:closed_s1} indicate that the initial value of $X(0)$ must be greater than $X^*$ by a factor $r \sim 1.5$.
Therefore, the value of $r$ in the condition (\ref{eq:closed_cd}) seems to take a similar value for $N=1$ and $N=3$. The value of $r$ is discussed in Appendix II.

\subsection{B. Preparation of critical conditions}

How the critical boundary condition or the initial condition expressed by the Eq.~(\ref{eq:open_cd}) and Eq.~(\ref{eq:closed_cd}) could have been prepared in the pre-RNA world for starting self-replication dynamics? The density of pn-molecules gradually increased by polymerization in some energy rich soup of mn-molecules. Then, in some open or closed networks, the geometric means of the density of pn-molecules $X$ might have had chance to satisfy the critical condition given by Eq.~(\ref{eq:open_cd}) or (\ref{eq:closed_cd}). In order to discuss the critical condition in more details, it is necessary to specify the form of catalytic reaction $P(n,i;n',i')$ of Eq.~(\ref{eq:dzn2}) which depends on the length $n$ and $n'$ and the information $i$ and $i'$ carried by interacting two molecules.

Although the dependence of $P(n,i;n',i')$ on $i$ and $i'$ is important for discussing evolutionary selection, we limit ourselves in this paper to discuss on the length dependence of the critical condition of self-replication for a closed network composed of pn-molecules of equal length $m$. It may be probable from the nature of catalytic interaction that the reaction constant may not explicitly depend on the length of molecules when the monomer is saturated as the source monomers. However, when the source density is not saturated, the catalytic interaction constant of two pn-molecules is considered to depend on the concentration $C$ of the monomers as shown in Section II, i.e., the dynamics (\ref{eq:dzt}) is replaced by Eq.~(\ref{eq:dzt_r}).

\begin{figure}
\begin{center}
  \includegraphics[height=6.0cm,keepaspectratio]{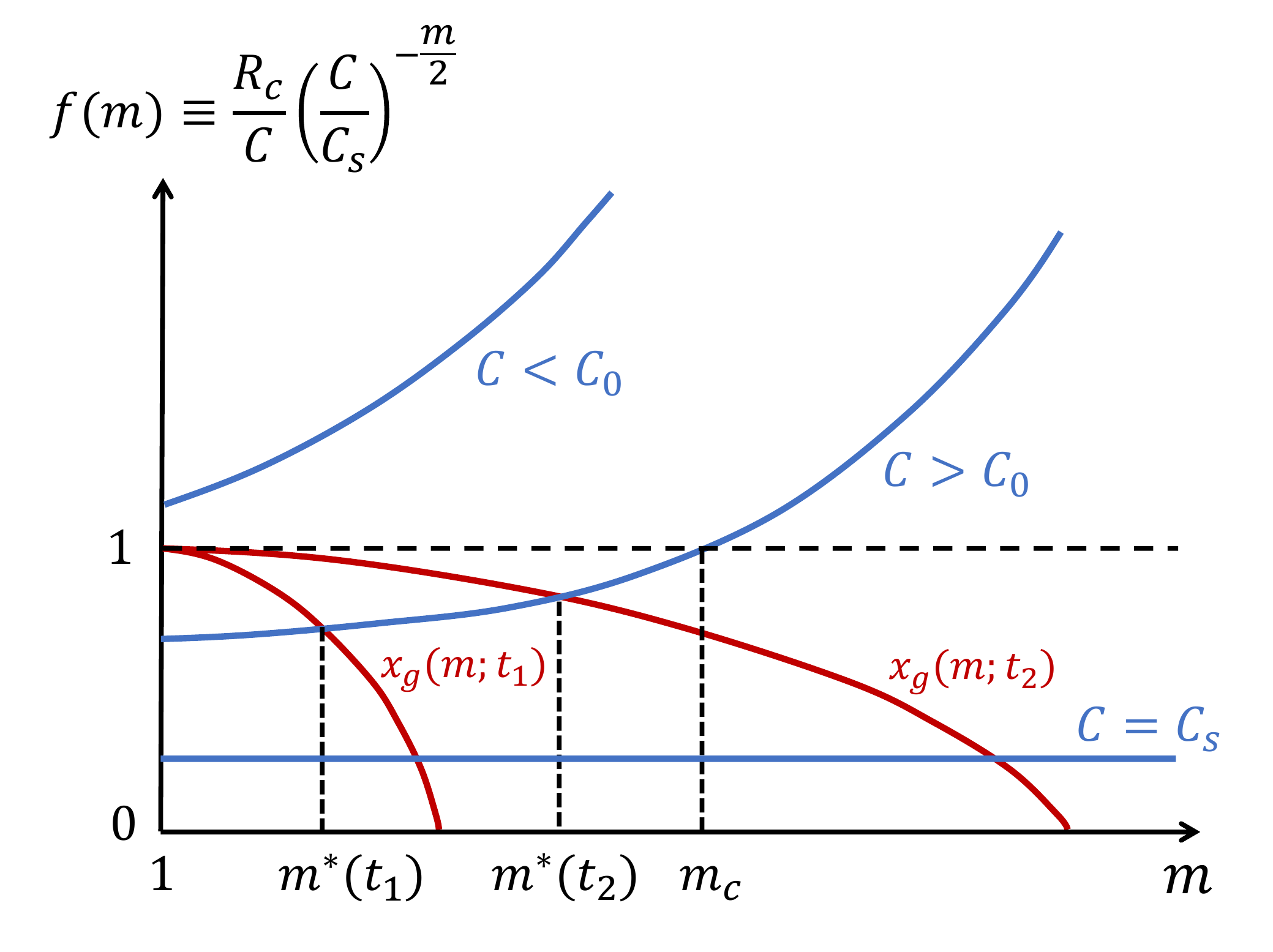}
  \caption{Condition for the onset of self-replicator of pn-molecules of length $m$. $x_g(m,t)$ and $f(m)$ are schematically drawn by the red and blue lines. $x_g$ grows with time $t$. There is no solution for $m$ satisfying the condition given by Eq.~(\ref{eq:onset2}), when $C < C_0$. When $C > C_0$ Eq.~(\ref{eq:onset2}) is satisfied for $m$ smaller than $m^*(t_1)$ and $m^*(t_2)$, for example, with a limitation $m < m_{\rm c}$. When $C = C_{\rm s}$ , $f(m)$ stays below unity and there is no limitation for the length $m$ of pn-molecules which satisfies Eq.~(\ref{eq:onset2}).
}
\label{fig:scenario}
\end{center}
\end{figure}

Then, by replacing $p_u$ by $p_u (C/C_s)^{m}$, the critical onset condition Eq.~(\ref{eq:closed_cd}) will be modified as
\begin{equation}
  X_g(m) > R_{\rm c} \left(\frac{C}{C_{\rm s}}\right)^{-m/2},
  \label{eq:onset}
\end{equation}
where
\begin{equation}
  R_{\rm c} \equiv \sqrt{2} r \left[\tau_z \tau_x \left(\prod_{u=1}^N p_u q_u \right)^{1/N} \right]^{-1/2}.
  \label{eq:Rc}
\end{equation}
Normalizing Eq.~(\ref{eq:onset}) by $C = X_g(1)$, Eq.~(\ref{eq:onset}) is written as
\begin{equation}
  x_g(m) > f(m),
  \label{eq:onset2}
\end{equation}
where
\begin{equation}
  x_g(m) \equiv \frac{X_g(m)}{C},~~~  f(m) \equiv \frac{R_{\rm c}}{C} \left(\frac{C}{C_{\rm s}}\right)^{-m/2}.
  \label{eq:fm}
\end{equation}  
The left-hand side of Eq.~(\ref{eq:onset2}) increases with time by polymerization and decreasing function of $m$, while right-hand side is fixed with time and increasing function of $m$, as shown in Fig.~\ref{fig:scenario}.
In order for the onset condition Eq.~(\ref{eq:onset2}) to be satisfied for some $m$, the right-hand side of Eq.~(\ref{eq:onset2}) for $m=1$ must be smaller than unity. This condition gives the value of the lowest concentration limit $C_0$ for the onset of self-replication of the shortest polymers as
\begin{equation}
  C_0 = (C_{\rm s} R_{\rm c}^2)^{1/3}.
  \label{eq:C0}
\end{equation}
From an obvious relation $C_{\rm s} > C_0$, the following relation is required for this discussion to be valid,
\begin{equation}
  C_{\rm s} > R_{\rm c}.
\end{equation}
As shown in Fig.~\ref{fig:scenario}, there is no solution for $m$ satisfying the condition given by Eq.~(\ref{eq:onset2}), when $C < C_0$. When $C > C_0$ Eq.~(\ref{eq:onset2}) is satisfied for $m$ smaller than $m^*(t_1)$ and $m^*(t_2)$, for example, with a limitation $m < m_{\rm c}$. When $C = C_{\rm s}$ , $f(m)$ stays below unity, and there is no limitation for the length m of pn-molecules which satisfies Eq.~(\ref{eq:onset2}).

The reality would have been more complex than Fig.~\ref{fig:scenario}. The length $m$ of the pn-molecules would have varied within a network. In this case, $x_g(m)$ and $(C/C_{\rm s})^{-m/2}$ in Eq.~(\ref{eq:fm}) have to be replaced by $x_g(\bar{m}, t)$ and $(C/C_{\rm s})^{-m_a/2}$ , where $\bar{m}$ is a statistical average of $m$, and $m_a$ is the algebraic mean of $m_j$ over the network. Normally $\bar{m}$ increases with $m_a$, and $x_g(\bar{m})$ will be decreasing function of $m_a$. And we can expect that the system will normally have a solution for the critical condition similar to $C=C_0$ for a special $m_a$, though further statistical research would be needed for this problem.

\subsection{C. Numerical values for the present model and comparison with 
  existing autocatalytic model}

The present model is based on the view point that the transition from material world to pre-RNA world existed at some material world time. Although some values $p$ and $q$ of chemical reaction and natural decay constants $\tau_x$  and $\tau_z$ have been reported in some chemical and thermodynamic conditions \citep{kiedrowski86,joice02,schrum10}, they might be far different from those in different prebiotic conditions. And one cannot at present claim the condition (\ref{eq:open_cd}) or (\ref{eq:closed_cd}) was satisfied or not at prebiotic time.

On the other hand, the existing autocatalytic model which is based on the assumption that the ligases had existed from the beginning of the material world \cite{kauffman93,kauffman86,mossel05,hordijk15,hordijk17,hordijk18}. It would be necessary to show in future how the ligase happened to appear in the early material world. The future research will clarify which model is closer to the reality of the mechanism to start pre-RNA world.

\subsection{D. Self-replication as a dissipative structure of a system far from thermodynamic equilibrium}

Another question apart from the detailed dynamics is what physical principle was associated with the onset of self-replication.

When a system is set far from equilibrium, the system has a potential to create `dissipative structures' \cite{glansdorff71} known as the dynamic structures in open systems, such as biological pattern formation \cite{shimizu83,sawai00} and fluid turbulence \cite{ozawa01}.
A dissipative structure is, in general, characterized by an internal current, which is set on when the degree how far the system is from thermal equilibrium exceeds a critical value. In case of fluid convection, a cyclic fluid current is set on in real space, when the Rayleigh number, the ratio of energy gain due to the buoyance force and the energy diffusion by thermal conduction and friction exceeds a critical value.

In the present self-replication problem, the saturation of the mononucleotide density at a value $C_0$ given by Eq.~(\ref{eq:C0}) satisfies a thermodynamic critical point far from equilibrium, similar to the Rayleigh number. At this point, the initial condition of the self-replication engine satisfies Eq.~(\ref{eq:closed_cd}), and the dynamics (\ref{eq:closed1}) and (\ref{eq:closed2}) is set on for increasing fluctuation towards self-replication. The cyclic internal chemical reaction current would contribute to increasing entropy production exponentially and stabilize dissipative structures in an environment far from thermodynamic equilibrium \cite{glansdorff71}.

This similarity of the self-replicator with fluid convection as dissipative structure was utilized in a laboratory experiment for the amplification of the self-replication of DNA by confining the molecules in a cell where a fluid convection was excited \cite{braun03}.  Although DNA self-replication is not mutually catalytic, the effective coupling of the two dissipative structures amplified the self-replication.

When we could identify the very early stage of self-replicator, which is the indispensable mechanism of life other than anything else, as a `dissipative structure', the answer to the question ``How life had physically started?'' might be ``It started by the second law of the thermodynamics in a prebiotic world when a dense assembly of mono-nucleotides achieved a state far from equilibrium.'' How the dissipative structures in general are created by the second law of thermodynamics or a lemma of it has been discussed elsewhere \cite{ziegler63,sawada81,schneider94,sawada20,kleidon04,martushev06}, and we do not go into the detail in this paper.

\section{VII. Conclusion}

A theoretical model for the onset of self-replication of informative molecules as a transition from a material world to the beginning of RNA world was presented. A quantitative expression of the condition for the self-replication of polynucleotide molecules towards self-replicators was obtained. In addition, the range of the length of the self-replication pn-polymers was theoretically predicted. The obtained results of the research would be helpful for designing future experiments for the self-replication of RNA molecules. The present research implies that the self-replication system belongs to the dissipative structures which are known to exist in a system far from thermodynamic equilibrium, and, therefore, that the initiation of life would be deeply connected to the second law of thermodynamics.


\appendix
\section{Appendix: Mode Selection Analysis of the Growth Pattern of Closed Networks of $N$ Self-replicator Units}

An analysis is shown here to demonstrate what mode can be selected among the excitation modes of a closed network with $N$ self-replicator units, and what mode shows the fastest exponential growth from the saddle point.

Eliminating $\delta Z_u$ from Eqs.~(\ref{eq:linearized1}) and (\ref{eq:linearized2}) we obtain,
\begin{align}
  &\tau_z\tau_x \frac{d^2 \delta X_u}{dt^2} + (\tau_z + \tau_x)\frac{d \delta X_u}{dt} - \tau_z \frac{d \delta X_{u-1}}{dt} \nonumber \\
  &~~~~~~~~~- (\delta X_{u-1} + \delta X_{u+1}) = 0,~~~ u=1,2,\dots N.
\end{align}
In order to obtain the eigenvalue of the dynamics, we set $\delta X_u \propto \exp(\lambda_k t + iku)$, where $i=\sqrt{-1}$ and $k=2\pi s/N ~~ (s=1,2,\dots N)$, as a cyclic boudary condition. Then, we obtain an equation for $\lambda_k$ as
\begin{equation}
  \tau_z \tau_x \lambda_k^2 + [(1-e^{-ik})\tau_z + \tau_x]\lambda_k - 2\cos k=0.
  \label{eq:lambda}
\end{equation}

Hereafter we simplify the analysis by setting $\tau_z = \tau_x = 1$. We do not lose the essential point of analysis by this simplification. Then Eq.~(\ref{eq:lambda}) is rewritten as
\begin{equation}
  \lambda_k^2 + (2-e^{-ik})\lambda_k - 2\cos k = 0.
  \label{eq:lambda2}
\end{equation}
We show the eigen-values $\lambda_k = \lambda_R(k) + i\lambda_I(k)$ for typical values of $k$, instead of general expression of $\lambda_k$, because they give a clearer view:
\begin{eqnarray}
  &k&=0, ~~~  (\lambda_R, \lambda_I) = (1,0)\; \& \;(-2,0), \nonumber\\
  &k&=\pi/2, ~~~ (\lambda_R, \lambda_I) = (0,0)\; \& \;(-2,-1), \nonumber\\
  &k&=\pi, ~~~ (\lambda_R, \lambda_I) = (-1,0)\; \& \;(-2,0), \nonumber\\
  &k&=3\pi/2, ~~~ (\lambda_R, \lambda_I) = (0,0)\; \& \;(-2,1).\nonumber
\end{eqnarray}
The only growth mode is at $k=0$. The mode with negative $\lambda_R$ decays with time, irrespective of $\lambda_I$. For modes near $k=0$, i.e., modes of small $k$, one can show from Eq.~(\ref{eq:lambda2}) that $\lambda_R(k) \approx 1-(31/54)k^2$ and $\lambda_I(k) \approx -k/3$. The results imply that the modes of small $k$ can grow with temporal oscillation. However, the fastest mode of growth is at $k=0$ without temporal oscillation.

\section{The Factor $r$ in the Critical Initial Condition}

We show here an analysis of the initial condition for the model used for Fig.~\ref{fig:closed_s1}.
\begin{eqnarray}
  \frac{dZ}{dt} = pX^2 -\frac{Z}{\tau_z}, ~~~ \frac{dX}{dt} = qZX - \frac{X}{\tau_x}.
\end{eqnarray}
By linearizing $X$ and $Z$ about the stationary point $(X^*, Z^*)$, $X = X^*+x$, $Z = Z^*+z$, where $X^* = (pq\tau_x \tau_z)^{-1/2}$, $Z^*=(q \tau_x)^{-1}$, we obtain
\begin{eqnarray}
  \frac{dz}{dt} = 2pX^*x-\frac{z}{\tau_z}, ~~~ \frac{dx}{dt} = qX^*z,
\end{eqnarray}
which gives us eigenvalues
\begin{eqnarray}
  \lambda_+ &= (1/2\tau_z)[(1+8\tau_z/\tau_x)^{1/2} -1], \nonumber\\
  \lambda_- &= -(1/2\tau_x)[(1+8\tau_z/\tau_x)^{1/2} +1], \nonumber
\end{eqnarray}
together with eigenvectors
\begin{eqnarray}
  (dz/dx)_+ &= (1/2)(p\tau_x/q\tau_z)^{1/2}[(1+8\tau_z/\tau_x)^{1/2} -1], \nonumber\\
  (dz/dx)_- &= -(1/2)(p\tau_x/q\tau_z)^{1/2}[(1+8\tau_z/\tau_x)^{1/2} +1]. \nonumber
\end{eqnarray}
The linear line with this gradient starting from general stationary point $(X^*, Z^*)$ cuts $Z=0$ line at
\begin{equation}
  X_0 = X^* - \frac{Z^*}{(dz/dx)_-} = X^*\left[1+\frac{2(\tau_z/\tau_x)^2}{(1+8\tau_z/\tau_x)^{1/2}+1}\right].
\end{equation}
For the case $\tau_z/\tau_x = 1$ of Fig.~\ref{fig:closed_s1}, one obtains $X_0 = (3/2) X^*$.

Fig.~\ref{fig:closed_s1} shows that the flow line of $N=1$ model starting from $(X=16, Z=0)$ grows in $X$ and $Z$ values for large $t$, while the flow line starting from $(X=15, Z=0)$ decays. The critical initial value lies between 15 and 16. The reason for this value is due to the fact that the negative eigenvalue of the dynamics at the stationary point $(10, 10)$ gives $(dz/dx)_- = -2$. The linear line with this gradient starting from $(10,10)$ cuts $Z=0$ line at $X_0=15$. The nonlinearity of the dynamics slightly modifies this line and cutting point is slightly higher than $X_0=15$ as described in the figure caption.

Numerical simulation for $N>1$, showed a similar behavior for the stationary value $X_g^*$ and the initial value $X_g(0)$ of the geometric mean of $X$.

\end{document}